\documentclass[conference]{IEEEtran}
\IEEEoverridecommandlockouts
\usepackage{cite}
\usepackage{amsmath,amssymb,amsfonts}
\usepackage{algorithmic}
\usepackage{graphicx}
\usepackage{textcomp}
\usepackage{xcolor}
\usepackage{threeparttable}
\usepackage{array}
\usepackage{booktabs}
\usepackage{multirow}
\usepackage{epsfig}
\usepackage{bm}
\usepackage{algorithm}
\usepackage{diagbox}
\usepackage{amssymb}
\usepackage{makecell}
\usepackage{adjustbox}
\usepackage{enumitem}
\usepackage{marvosym}
\usepackage{braket}
\usepackage{color}
\def\BibTeX{{\rm B\kern-.05em{\sc i\kern-.025em b}\kern-.08em
    T\kern-.1667em\lower.7ex\hbox{E}\kern-.125emX}}
\begin{document}

\title{An Experimental Study on Joint Modeling for Sound Event Localization and Detection with Source Distance Estimation}

\author{
    \IEEEauthorblockN{
        Yuxuan Dong,
        Qing Wang\IEEEauthorrefmark{1},\thanks{\IEEEauthorrefmark{1}Corresponding author}
        Hengyi Hong,
        Ya Jiang,
        Shi Cheng
    }
    \IEEEauthorblockA{NERC-SLIP, University of Science and Technology of China (USTC), Hefei, China}
    \IEEEauthorblockA{\{yxdong0320, hyhong, yajiang, chengshi\}@mail.ustc.edu.cn, qingwang2@ustc.edu.cn}
}

\maketitle
%\vspace{-5pt}
\begin{abstract}
In traditional sound event localization and detection (SELD) tasks, the focus is typically on sound event detection (SED) and direction-of-arrival (DOA) estimation, but they fall short of providing full spatial information about the sound source. The 3D SELD task addresses this limitation by integrating source distance estimation (SDE), allowing for complete spatial localization. We propose three approaches to tackle this challenge: a novel method with independent training and joint prediction, which firstly treats DOA and distance estimation as separate tasks and then combines them to solve 3D SELD; a dual-branch representation with source Cartesian coordinate used for simultaneous DOA and distance estimation; and a three-branch structure that jointly models SED, DOA, and SDE within a unified framework.  Our proposed method ranked first in the DCASE 2024 Challenge Task 3, demonstrating the effectiveness of joint modeling for addressing the 3D SELD task. The relevant code for this paper will be open-sourced in the future.
%Validation on the STARSS23 dataset shows that our system effectively handles the 3D SELD task.
\end{abstract}


\section{Introduction}
Sound event localization and detection (SELD) aims to detect the temporal activity of sound events and track their spatial locations using multichannel audio inputs. The spatio-temporal data provided by SELD systems is useful for various machine awareness applications, including smart homes and audio surveillance\cite{foggia2015audio,valenzise2007scream}.

3D SELD consists of three subtasks\cite{krause2024sound}: sound event detection (SED), sound source localization (SSL), and source distance estimation (SDE). SED uses classification methods such as Gaussian mixture models (GMM)\cite{heittola2013context}, hidden Markov models (HMM)\cite{butko2011two}, and recurrent neural networks (RNN)\cite{hayashi2017duration} to identify the time occurrence of sound events. SSL estimates the direction-of-arrival (DOA) using both traditional\cite{schmidt1986multiple,roy1989esprit} and deep learning-based methods\cite{adavanne2018direction}. Research on SDE has mainly focused on binaural formats, though \cite{kushwaha2023sound} expanded this by exploring loss functions and integrating activity detection with a tetrahedral microphone array for distance estimation.

Development of complex scene analysis systems has driven the emergence of multi-task learning methods comparing to single-task models. In \cite{adavanne2018sound}, a CRNN model with dual-branch output was proposed, jointly modeling the SED and DOA tasks. Later, activity-coupled Cartesian DOA (ACCDOA)\cite{shimada2021accdoa} and multi-ACCDOA\cite{Shimada2022} representations improved the model structure by eliminating the need to balance dual-branch objectives. Since then, numerous models have been proposed to better handle the SELD task in complex scenarios. The channel-spectro-temporal transformer (CST-Former)\cite{shul2024cst} enhanced performance by applying attention mechanisms across time, frequency, and channel domains. Models like the event independent network (EIN)\cite{cao2020event,hu2022track} and angular-distance-based SELD (AD-YOLO)\cite{kim2023ad} improved the handling of polyphonic problems. Additionally, the multi-modal SELD method in \cite{yasuda20246dof} accounted for microphone self-motion, using both audio and motion-tracking sensor data for movable human.

In \cite{krause2023binaural}, the authors explored a joint modeling manner for binaural SDE and DOA estimation using motion cues. The first study to jointly model all three tasks, namely detecting sound categories, estimating DOA and distance of sources, in 3D SELD was presented in \cite{krause2024sound}, introducing two approaches: a multi-task method with separate branches for SELD and SDE, and an extended multi-ACCDOA method that incorporates distance estimation into the vector.

3D SELD was introduced for the first time in the Detection and Classification of Acoustic Scenes and Events (DCASE) 2024 Challenge Task 3 \cite{diazbaseline}. There is rarely relevant research to handle this complex task. By carefully designing learning objectives, this paper focuses on exploring the impact of different output representation formats on the 3D SELD task. In this paper, we propose three methods for jointly modeling the 3D SELD task:

%However, current approaches to joint modeling in 3D SELD face challenges. The multi-task approach struggles with balancing the losses across the three branches, and further research is needed on the design of loss functions. Meanwhile, the extended multi-ACCDOA method increases model complexity, which can impair learning performance. In this paper, we propose three methods for jointly modeling the 3D SELD task:

\begin{itemize}
    \item We propose a novel method with independent training and joint prediction, that employs two models namely SED-DOA and SED-SDE, combining their outputs to gather the information needed for the 3D SELD task. As the joint modeling of SED and SDE is a new research area, we investigate the effects of different loss functions.
    \item We integrate the information from the DOA and SDE branches to obtain the spatial coordinates of sound events, simplifying the network output to a dual-branch representation for SED and source coordinate estimation (SCE) simultaneously.
    \item We explore a network with three branch outputs for SED, DOA, and SDE simultaneously, and compare the advantages and limitations of these approaches against the multi-ACCDOA representation.
\end{itemize} %[label=(\arabic*)]

\section{Method}
\begin{figure}
    \centering
    \includegraphics[width=1\linewidth]{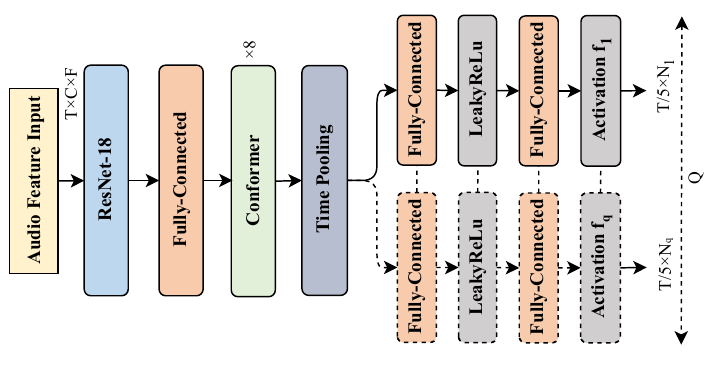}
    \caption{The proposed 3D SELD model. $Q$ and $N_{q}$ represent the number of output branches and the output dimensions of each branch.}
    \label{fig:model}
\end{figure}

The overall framework of our proposed 3D SELD model is illustrated in Fig.~\ref{fig:model}. The model input consists of 4-channel first-order Ambisonics (FOA) audio data, segmented into 10-second intervals. A short-time Fourier transform (STFT) with a 40 ms frame length and a 20 ms hop lentgh is applied to extract 4-channel log-Mel spectrograms and 3-channel intensity vector (IV) features. The resulting feature matrix, shaped as $T \times CH \times F$, is then fed into the model, where $T$, $CH$, and $F$ denote the temporal length, number of channels, and feature dimensions, respectively.

For an audio recording with $C$ classes and $T$ frames, the learning target of the 3D SELD is defined as $\mathbf{O}_{ct} = [a_{ct},\mathbf{R}_{ct},d_{ct}]$, where $c$, $t$ indicate the class index and frame index. $a_{ct} \in \{0,1\}$ stands for the event detection activity. $\mathbf{R}_{ct}=(x_{ct},y_{ct},z_{ct}) \in\langle-1,1\rangle$ denotes the DOA vector of sound event. $d_{ct}\in[\,0,\infty\rangle$ corresponds to source distance. $a_{ct}=1$ indicates that the event class $c$ is active at time $t$, in which case $\|\mathbf{R}_{ct}\|=1$ and $d$ is a positive value.
%\mathbf{R}_{ct}=(x_{ct},y_{ct},z_{ct}) \in\langle-1,1\rangle$ denotes the DOA vector of sound event, and $D_{ct}\in\langle0,\infty\rangle$ corresponds to source distance. For the active class $c'$, its DOA is a unit vector, i.e., $\|\mathbf{R}_{c't}\|=1$.  

The main network we use is the ResNet-Conformer (RC) architecture, which was introduced in the DCASE 2022 Challenge Task 3\cite{niu2023experimental}. It consists of an 18-layer ResNet and an 8-layer Conformer, with a time pooling layer applied after the Conformer. The model's output consists of $Q$ branches, where $Q$ varies depending on the specific task to be solved. Each branch includes two fully-connected (FC) layers with distinct activation functions. The following sections provide further details on each joint modeling method.

\subsection{Joint Modeling of SED-DOA and SED-SDE}
\subsubsection{Exploration on SDE branch}
In previous SELD research, mean squared error (MSE) was the most commonly used loss function for DOA estimation. However, incorporating distance estimation introduces a different value range, which may require alternative loss functions better suited to this task. Building on the findings from \cite{kushwaha2023sound}\cite{krause2024sound}, we evaluate the performance of three loss functions, namely MSE, mean square percent error (MSPE), and mean absolute percent error (MAPE) for optimizing the SDE branch. The loss for the SDE task $\mathcal{L}_{\mathrm{SDE}}$ can be expressed as follows:

\begin{align}
    MSE=\, &\frac{1}{CT}\sum_{c,t}||a_{ct}(d_{ct}-\hat{d}_{ct})||^{2},\\
    MSPE=\,&\frac{1}{CT}\sum_{c,t}||a_{ct}(d_{ct}-\hat{d}_{ct})/d_{ct}||^{2}, \\
    MAPE=\,&\frac{1}{CT}\sum_{c,t}||a_{ct}(d_{ct}-\hat{d}_{ct})/d_{ct}||,
\end{align}
where $d_{ct}$ and $\hat{d}_{ct}$ represent the ground truth and predicted distance, $a_{ct}$ represents the ground truth of event activity.

\subsubsection{Joint modeling of DOA and distance}
We propose a novel method with independent training and joint prediction to address the 3D SELD task. Firstly, we train two models, namely SED-DOA and SED-SDE, to estimate DOA and distance separately. SED-DOA and SED-SDE each output $Q=2$ branches. For SED-DOA, $N_1 = C$ and $N_2 = 3C$, while for SED-SDE, $N_1 = C$ and $N_2 = C$. The SED, DOA, and SDE branches use Sigmoid, Tanh, and ReLU activation functions, respectively. Both networks are trained with a multi-objective learning framework:
\begin{align}
    \mathcal{L}_{\mathrm{SED{\rule[0.3ex]{0.2em}{0.3pt}}DOA}}\!=\,&\beta_1\mathcal{L}_{\mathrm{SED}}+\beta_2\mathcal{L}_{\mathrm{DOA}},  \\ 
    \mathcal{L}_{\mathrm{SED{\rule[0.3ex]{0.2em}{0.3pt}}SDE}}\!=\,&\gamma_1\mathcal{L}_{\mathrm{SED}}+\gamma_2\mathcal{L}_{\mathrm{SDE}},
\end{align}
where $\beta_1=0.1$ and $\beta_2=1$ are used as the weights for SED loss and DOA loss according to our previous work \cite{jiang2024exploring}, $\gamma_1 $ and $\gamma_2$ represent the weights for SED loss and SDE loss, respectively. %Their impact on performance is discussed in the experiments. 
The SDE branch uses the loss functions shown in (1-3). Binary cross entropy (BCE) and MSE are used as the loss functions for the SED and DOA branches, respectively, and are defined as follows:
%\begin{equation}
%\mathcal{L}_{\mathrm{SED}}=-\frac{1}{C T} \sum_{c, t}[a_{ct} %\log \hat{a}_{ct} +(1-a_{ct}) \log (1-\hat{a}_{ct})],
%\end{equation}
% \nonumber \\[1mm]
%\begin{equation}
%\mathcal{L}_{\mathrm{DOA}}=\frac{1}{C T} \sum_{c, t}||a_{ct}%(\hat{\mathbf{R}}_{ct}-\mathbf{R}_{ct})||^2,
%\end{equation}
\begin{align}
\mathcal{L}_{\mathrm{SED}}\!=&\!-\!\frac{1}{C T} \!\sum_{c, t}[a_{ct} \log \hat{a}_{ct} +(1-a_{ct}) \log (1-\hat{a}_{ct})], \\
\mathcal{L}_{\mathrm{DOA}}\!=&\frac{1}{C T} \sum_{c, t}||a_{ct}(\hat{\mathbf{R}}_{ct}-\mathbf{R}_{ct})||^2,
\end{align}
where $a_{ct}$ and $\hat{a}_{ct}$ represent the ground truth and predicted event activity respectively, $\mathbf{R}_{ct}$ and $\hat{\mathbf{R}}_{ct}$ represent the ground truth and predicted DOA vectors respectively. 

During inference, predictions from both models are combined to yield the results required for 3D SELD. Let the SED outputs of both models be $\hat{a}^{(1)}_{ct}$ and $\hat{a}^{(2)}_{ct}$, the final prediction result is $\hat{\mathbf{O}}_{ct} = [\tilde{a}_{ct},\hat{\mathbf{R}}_{ct},\hat{d}_{ct}]$, where $\tilde{a}_{ct}=(\hat{a}^{(1)}_{ct}+\hat{a}^{(2)}_{ct})/2$ is computed from both models, $\hat{\mathbf{R}}_{ct}$ and $\hat{d}_{ct}$ are predicted from the SED-DOA and SED-SDE models, respectively. 

%\begin{align}
%    \hat{a}_{ct}=&h[\lambda \hat{p}_{1ct}+(1-\lambda)\hat{p}_{2ct}]  \\ 
%    h(x)=&\begin{cases}1&\mathrm{if}\: x\geq0.5\\0&\mathrm{if}\: x<0.5.\end{cases}
%\end{align}
%Here, $p$ represents the posterior probability of event activation, $h(x)$ is a threshold decision function, and $\lambda$ is a weighting factor that controls the influence of the SED results from the SED-DOA model on the final 3D SELD output.

\subsection{SED-SCE Model}
Since the DOA represents a unit direction vector, multiplying it by the distance gives the spatial Cartesian coordinates of the sound source, expressed as $\mathbf{S}_{ct} = d_{ct} \cdot \mathbf{R}_{ct}$. Based on this, we design a network with $Q = 2$ output branches, where $N_1 = C$ is used for SED and $N_2 = 3C$ is used for SCE. Given the difference in value ranges between the true Cartesian coordinates and the unit vectors, linear output is used for the SCE branch to capture the full ranges of true coordinates.

In this model, the overall loss function is expressed as:
\begin{equation}
% \label{eq: sed_doa_sde_loss}
\mathcal{L}_{\mathrm{SED{\rule[0.3ex]{0.2em}{0.3pt}}SCE}}=\eta_1\mathcal{L}_{\mathrm{SED}}+\eta_2\mathcal{L}_{\mathrm{SCE}},
\end{equation}
where $\eta_1$ and $\eta_2$ are the loss weights for the SED and SCE branches, respectively. The SED branch uses the same BCE loss function as in the previous section, while the SCE branch uses the MSE loss function, expressed as:
\begin{equation}
\mathcal{L}_{\mathrm{SCE}}= \frac{1}{C T} \sum_{c, t}||a_{ct}(\hat{\mathbf{S}}_{ct}-\mathbf{S}_{ct})||^2,
\end{equation}
where $\mathbf{S}_{ct}$ and $\hat{\mathbf{S}}_{ct}$ represent the ground truth and predicted source Cartesian coordinates, respectively. The estimated DOA unit vector and distance are obtained by $\hat{\mathbf{S}}_{ct}/||\hat{\mathbf{S}}_{ct}||$ and $||\hat{\mathbf{S}}_{ct}||$. 
%And the length of the DOA vector can be used to represent the source distance, expressed as $\hat{d}_{ct} = ||\hat{\mathbf{S}}_{ct}||$.
%\begin{align}
%    \hat{\mathbf{R}}_{ct} =& \frac{\mathbf{S}_{ct}}{\left\|\mathbf{S}_{ct}\right\|} \\
%    \hat{d}_{ct} =& \left\|\mathbf{S}_{ct}\right\|
%

\subsection{SED-DOA-SDE Model}
Another joint modeling method is to extend our proposed dual-branch model \cite{niu2023experimental} by adding an additional branch for distance estimation, resulting in $Q = 3$ output branches. Therefore, $N_1 = C$, $N_2 = 3C$, and $N_3 = C$ denote the number of nodes used for the SED, DOA, and SDE tasks, respectively. This method is trained with a multi-task learning framework, with the overall loss function defined as:
\begin{equation}
% \label{eq: sed_doa_sde_loss}
\mathcal{L}_{\mathrm{SED{\rule[0.3ex]{0.2em}{0.3pt}}DOA{\rule[0.3ex]{0.2em}{0.3pt}}SDE}}=\lambda_1\mathcal{L}_{\mathrm{SED}}+\lambda_2\mathcal{L}_{\mathrm{DOA}}+\lambda_3\mathcal{L}_{\mathrm{SDE}}.
\end{equation}
Here $\lambda_1$, $\lambda_2$ and $\lambda_3$ are the weights for the SED, DOA and SDE losses, respectively. 

Table~\ref{tab:parameters} summarizes the output parameters of the models mentioned above and provides a comparison with the parameters of the multi-ACCDOA model. The multi-ACCDOA model includes 3 tracks, and the 3-element DOA vector is extended to 4 to include the distance estimate\cite{krause2024sound}. Thus, $N_1= 3\times4C$. 

All models are implemented in PyTorch and optimized using the Adam optimizer \cite{kingma2014adam}. A three-stage learning rate scheduler \cite{park2019specaugment} is applied during training. The key training parameters include a batch size of 32, a total of 360k steps (with 100k steps used for the SED-SDE model), and an upper limit learning rate of 0.001.
\begin{table}
    \centering
    \caption{Output parameters for the proposed models.}
    \begin{tabular}{cccc}
        \toprule
         Models& Q & Nq & Activation Function\\
         \midrule
         Multi-ACCDOA\cite{krause2024sound} & 1 & 3$\times$4C & Linear\\
         SED-DOA & 2 & [C,3C] & [Sigmoid, Tanh]\\
         SED-SDE & 2 & [C,C] & [Sigmoid, ReLu]\\
         SED-SCE & 2 & [C,3C] & [Sigmoid, Linear]\\
         SED-DOA-SDE& 3 & [C,3C,C] & [Sigmoid,Tanh,ReLu] \\
         \bottomrule
    \end{tabular}
    \label{tab:parameters}
\end{table}

\section{Experiments}
\subsection{Data}
In our experiments, we use the audio data in FOA format of the STARSS23 dataset\cite{shimada2024starss23}, which contains 7 hours and 22 minutes of real recordings, split into training data (90 clips) and test data (78 clips). To expand the dataset, we applied two data augmentation methods. The first method involved simulating new multi-channel data using selected sound samples from the FSD50K dataset\cite{fonseca2021fsd50k} and the provided spatial room impulse responses (SRIRs) \cite{politis2020dataset}. This yielded approximately 40 hours of synthesized data using the Spatial Scaler library\cite{roman2024spatial}. The second method, audio channel swapping (ACS), is a spatial augmentation technique proposed in \cite{wang2023four}. It modifies audio channels based on the physical and rotational properties of spherical microphone arrays to increase DOA representation. However, since the ACS augmentation method does not enhance the diversity of source distance estimation, it is excluded during training the SED-SDE model.

\subsection{Evaluation Metrics}
The evaluation metric is an improvement version of the metric described in \cite{mesaros2019joint}. It assesses sound event detection and localization using the F1-score (F1) and DOA error (DOAE), while distance estimation performance is evaluated using the relative distance error (RDE). F1 is computed with two separate thresholds: an angular threshold of $20^\circ$ for DOAE and a relative distance threshold of 100\% for RDE. This approach allows distinct penalties for angular and distance estimation performance. Additionally, the SELD$_{score}$ here is calculated on frame-based evaluation rather than segment-based evaluation. The formula is as follows:
\begin{equation}
    \text{SELD}_{score}=\frac{1}{3}[(1-\text{F1})+\text{DOAE}/{180^\circ}+\text{RDE}].
\end{equation}

For the SED-DOA and SED-SDE models, the evaluation metric is slightly modified: the relative distance threshold restriction is omitted for the SED-DOA model, and the angular threshold restriction is omitted for the SED-SDE model. The SED-SDE$_{score}$ is defined by the following formula:
\begin{equation}
    \text{SED-SDE}_{score}=\frac{1}{2}[(1-\text{F1})+\text{RDE}].
\end{equation}

\subsection{Results and Analysis}
\subsubsection{Ablation studies on SDE branch}

We began with a dual-branch SED-SDE model and investigated how different loss functions and weights affect the results. We explored MSE, MSPE, and MAPE loss functions in the SDE branch. $[\gamma_1, \gamma_2]=[0.1,1]$ indicates that the loss weights for the two branches are set to 0.1 and 1, respectively.
We recorded the loss values of both branches during training and validation, finding that the ratio of $\mathcal{L}_{\mathrm{SED}}$$/$$\mathcal{L}_{\mathrm{SDE}}$ was approximately 10. To balance the losses between the two tasks, we maintained $\gamma_2 / \gamma_1$ a similar order of magnitude and conducted a series of ablation experiments. Table \ref{tab:my_label2} shows the experimental results.

As $\gamma_2 / \gamma_1$ increases, the RDE performance improves for all three loss functions, aligning with our expectation that increasing the SDE branch's loss weight enhances distance estimation. However, we observed that an excessively small loss weight for the SED branch may lead to gradient vanishing problem, causing a sharp decline in F1 scores. This is because when the actual SED loss used to update the network parameters, i.e., $\gamma_1 \mathcal{L}_{\mathrm{SED}}$, becomes relatively small compared to $\gamma_2 \mathcal{L}_{\mathrm{SDE}}$, the SED task cannot be effectively learned. Ultimately, the best performance for SED was achieved using the MSE loss for the SDE branch, with equal loss weights of $[1,1]$. Conversely, the MSPE loss function with loss weights of $[0.1,2]$ yielded the best performance for SDE.

\begin{table}[t]
    \centering
    \caption{Ablation studies with loss functions and weighs for the SED-SDE model on the development set of STARSS23 dataset.}
    \label{tab:my_label2}
    \begin{tabular}{c|c|c|c|c}
    \toprule
         $\mathcal{L}_{\mathrm{SDE}}$ & [$\gamma_1$, $\gamma_2$]  & F1 $\uparrow$ & RDE $\downarrow $& SED-SDE$_{score}$ $\downarrow$\\
         \midrule
         \multirow{3}{*}{MSE} &[1,1] & \textbf{0.62} & 0.26 & 0.320 \\
        &[0.1,1]& 0.50  & 0.26 & 0.380\\
        &[0.1,2]& 0.37  & 0.25 & 0.440\\
        \midrule
        \multirow{3}{*}{MAPE} &[1,1] & 0.58 & 0.29 & 0.355\\
        &[0.1,1]& 0.57 & 0.26 & 0.345\\
        &[0.1,2]& 0.55 & 0.24 & 0.345\\
        \midrule
        \multirow{3}{*}{MSPE} &[1,1] & 0.59  & 0.29 & 0.350\\
        &[0.1,1]& 0.58 & 0.26 & 0.340\\
        &[0.1,2]& 0.57  & \textbf{0.23} & 0.330\\
    \bottomrule
    \end{tabular}
\end{table}

\subsubsection{Results on joint modeling methods}
\begin{table}
    \centering
    \caption{Performances comparison among different joint modeling methods on the development set of STARSS23 dataset}
    \label{tab:my_label3}
    \begin{tabular}{c|c|c|c|c}
    \toprule
         Methods & F1 $\uparrow$ & DOAE $\downarrow$ & RDE $\downarrow$ & SELD$_{score}$ $\downarrow$ \\
         \midrule
         Multi-ACCDOA\cite{krause2024sound} & 0.44 & 16.7$^{\circ}$ & 0.32 & 0.324 \\
         SED-DOA& 0.53 & 14.6$^{\circ}$ & -  & - \\
         SED-SDE&  0.57 & - & 0.23 & -\\
         SED-SCE& 0.46 & 15.3$^{\circ}$ & 0.26 & 0.295\\
         SED-DOA-SDE& 0.45 & 15.6$^{\circ}$ & 0.27 & 0.302\\
         SED-DOA + SED-SDE& \textbf{0.53} & \textbf{14.6$^{\circ}$} & \textbf{0.23} & \textbf{0.260}\\
         \bottomrule
    \end{tabular}
\end{table}

Table \ref{tab:my_label3} presents the performance of the three proposed joint modeling methods on the STARSS23 development dataset, compared to the existing multi-ACCDOA approach \cite{krause2024sound}. `Multi-ACCDOA' refers to a RC model (the same as used in our methods) combined with the multi-ACCDOA representation. The SED-SDE model listed uses the MSPE loss function for the SDE branch with loss weights set to $[0.1, 2]$. Although this model does not achieve the best SED-SDE$_{score}$, it is chosen for providing the best SDE performance, which is the primary objective in designing this model structure. 

For the SED-SCE and SED-DOA-SDE models, the loss functions and weights were optimized to the best-known values. The SED-SCE model uses MSE for the SCE branch with loss weights of $[\eta_1,\eta_2]=[1,1]$, while the SED-DOA-SDE model uses MSPE for the SDE branch with loss weights of $[\lambda_1,\lambda_2,\lambda_3]=[0.1, 1, 2]$.

All three proposed methods outperform the multi-ACCDOA approach \cite{krause2024sound}, demonstrating their effectiveness in jointly modeling the three subtasks in 3D SELD. The SED-SCE and SED-DOA-SDE methods achieve similar results, with the SED-SCE model showing a relative improvement of 9.0\% over the multi-ACCDOA method in the overall score. When using the proposed  method with independent training and joint prediction, we observed the best 3D SELD performance across all three evaluation metrics shown in the bottom row of Table~\ref{tab:my_label3}. Compared to the multi-ACCDOA method, our joint modeling approach achieved a relative improvement of 20.5\% in F1, 12.5\% in DOAE, and 28.1\% in RDE.

\subsubsection{Comparison with other methods}
\begin{table}[h]
    \centering
    \caption{Performances comparison with other methods on the development set of DCASE 2024 Challenge Task 3.}
    \label{tab:my_label4}
    \begin{tabular}{c|c|c|c|c}
    \toprule
         System  &F1 $\uparrow$ & DOAE $\downarrow$ & RDE $\downarrow$  & SELD$_{score}$ $\downarrow$ \\
         \midrule
         CRNN [3] & 0.13 & 36.9$^{\circ}$  & 0.33 &0.468 \\
         \midrule
         RC-AFF [31] & 0.44 & 13.7$^{\circ}$  & 0.30 &0.312\\
         RC-SE [32]  & 0.34 & 20.4$^{\circ}$  & 0.30 &0.358\\
         CST-Former [33]  & 0.35 & 18.8$^{\circ}$  & 0.28 &0.345 \\
         Proposed  &\textbf{0.40}  & \textbf{19.4$^{\circ}$} & \textbf{0.23} &\textbf{0.313}\\
         \midrule
         SED-DOA + SED-SDE &\textbf{0.59}  & \textbf{12.9$^{\circ}$} & \textbf{0.23} &\textbf{0.237}\\
        \bottomrule
    \end{tabular}
\end{table}

We compare our proposed joint modeling methods with several top-ranked systems of DCASE 2024 Challenge Task 3. Table~\ref{tab:my_label4} presents the results on the development set. 
`CRNN' refers to the baseline system in \cite{krause2024sound}. `RC-AFF', `RC-SE' and `CST-Former' ranked 4th, 3rd, and 2nd, respectively, in the evaluation set of DCASE 2024 Challenge Task 3. `RC-AFF' refers a ResNet-Conformer architecture with attention feature fusion (AFF) incorporated into the ResNet block \cite{Guan_CQUPT_task3_report}. `RC-SE' represents the method in \cite{Yeow_NTU_task3a_report}, where the ResNet-Conformer model is combined with spatial and channel Squeeze-and-Excite (SE) blocks. `CST-Former' is a transformer-based network incorporated with DOA and event guidance attention blocks \cite{Yu_HYUNDAI_task3a_report}. Notably, the baseline and all three of these other approaches utilize the multi-ACCDOA output representation. Both `RC-AFF' and `RC-SE' used the ACS method to increase the training size, while `CST-Former' was trained without the ACS method.

Our system (denoted as `SED-DOA + SED-SDE') achieves a SELD$_{score}$ of 0.237 without model ensemble, ranking first place. It was trained using a pre-trained model as a starting point, with details provided in our technical report \cite{Du_NERCSLIP_task3_report}. The proposed joint modeling method outperforms all four systems by a large margin when training with a large data size. For a fair comparison, we train a model using the same amount of data as CST-Former \cite{Yu_HYUNDAI_task3a_report}, denoted as `Proposed'. Compared to CST-Former, our model achieved better overall metric, with a slightly lower DOAE but significantly higher F1 and RDE scores. This underscores the effectiveness of the method with independent training and joint prediction.

\section{Conclusion}
In this study, we explored three joint modeling approaches for 3D SELD task and investigated suitable loss functions for SDE. We propose a novel method featuring independent training and joint prediction, treating DOA and SDE as separate tasks for better modeling. Additionally, we introduce two multi-task learning models to address the complex 3D SELD task within a unified framework. The proposed joint modeling methods achieve the best performance for 3D SELD, ranking first place in DCASE 2024 Challenge Task 3. 

\section*{Acknowledgment}
This work was supported by the National Natural Science Foundation of China under Grant No. 62401533.

%\vfill
%\pagebreak
\bibliographystyle{IEEEtran}
\bibliography{refs}

\end{document}